\documentclass{article}
\usepackage{ijcai}

\usepackage{amssymb}
\usepackage{amsmath}
\usepackage{amsthm}
\usepackage{booktabs}
\usepackage{graphicx}
\usepackage{multirow}
\usepackage{tabularx}
\usepackage{times}
\usepackage{url}
\usepackage[small]{caption}
\usepackage[hidelinks]{hyperref}

\newcommand{\ie}{\textit{i.e.,} }
\newcommand{\eg}{\textit{e.g.,} }
\newcommand{\etal}{\textit{et al.} }
\newcommand{\email}[2]{\href{mailto:#2}{#1}}

\title{End-to-End Personalized Next Location Recommendation via \\ Contrastive User Preference Modeling}

\author{
Yan~Luo$^{1,2}$ \and Ye~Liu$^1$\thanks{Corresponding Author} \and Fu-lai~Chung$^1$ \and Yu~Liu$^3$ \and Chang~Wen~Chen$^1$
\affiliations
$^1$Department of Computing, The Hong Kong Polytechnic University \\
$^2$Media Lab, Massachusetts Institute of Technology\\
$^3$School of Earth and Space Sciences, Peking University
\emails
\{\email{silver.luo}{silver.luo@connect.polyu.hk},\email{coco.ye.liu}{coco.ye.liu@connect.polyu.hk}\}@connect.polyu.hk,
\email{cskchung@comp.polyu.edu.hk}{cskchung@comp.polyu.edu.hk} \\
\email{liuyu@urban.pku.edu.cn}{liuyu@urban.pku.edu.cn},
\email{changwen.chen@polyu.edu.hk}{changwen.chen@polyu.edu.hk}}

\begin{document}

\maketitle

\begin{abstract}
Predicting the next location is a highly valuable and common need in many location-based services such as destination prediction and route planning. The goal of next location recommendation is to predict the next point-of-interest a user might go to based on the user's historical trajectory. Most existing models learn mobility patterns merely from users' historical check-in sequences while overlooking the significance of user preference modeling. In this work, a novel Point-of-Interest Transformer (POIFormer) with contrastive user preference modeling is developed for end-to-end next location recommendation. This model consists of three major modules: history encoder, query generator, and preference decoder. History encoder is designed to model mobility patterns from historical check-in sequences, while query generator explicitly learns user preferences to generate user-specific intention queries. Finally, preference decoder combines the intention queries and historical information to predict the user's next location. Extensive comparisons with representative schemes and ablation studies on four real-world datasets demonstrate the effectiveness and superiority of the proposed scheme under various settings.
\end{abstract}

\section{Introduction}

Geo-tagged data has already become a major and common data source that is readily available in daily lives. The proliferation of location-based services (LBS) such as Gowalla and Foursquare has attracted users to share their visited locations and post comments on point-of-interests (POIs). In this context, next location recommendation, which exploits a user's historical check-in sequences to predict next POI the person may visit, is becoming increasingly important to improve user experience in LBS. Location recommender systems, on the one hand, can be employed to achieve personalized route planning for end-users. On the other hand, it also has huge potential commercial value in advertisements. Therefore, next location recommendation has attracted extensive attention from both academia and industry.

\begin{figure}
\centering
\vspace{0.5mm}
\includegraphics[width=0.9\linewidth]{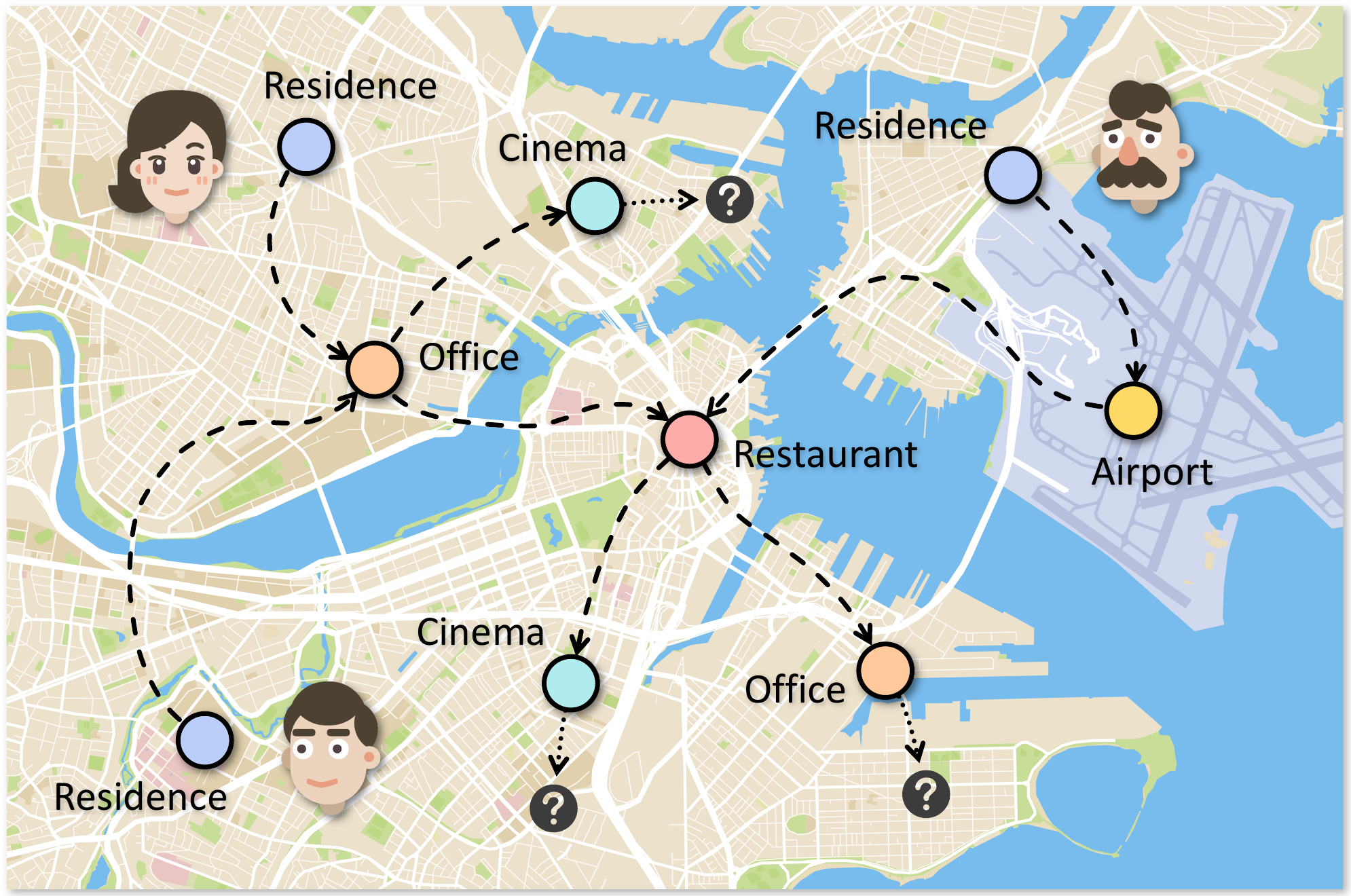}
\caption{Illustration of the next location recommendation task. Dashed lines indicate historical trajectories. Our goal is to predict the next location that a user might go to (denoted by dotted lines) based on the historical check-ins.}
\label{fig:1}
\end{figure}

In recent years, how to enhance the performance of next location recommendation has been extensively studied \cite{hang2018exploring,li2018next,zhao2020where,afzali2021pointrec}. Figure~\ref{fig:1} shows the illustration of next location recommendation. It is worth noting that location recommendation is essentially an application of sequential recommendation. Several previous studies \cite{zhang2015spatiotemporal,wang2018tpm} demonstrated that successive check-ins in a sequence are usually correlated. Statistical models such as Markov Chain \cite{gambs2012next,baumann2013influence,liu2013personalized} have been widely employed to solve this problem at the beginning. Subsequently, with the development of deep learning, RNN-based methods \cite{liu2016predicting,zhu2017next,feng2018deepmove} and graph-based methods \cite{xie2016learning,wang2022graph} have been developed. Recently, researchers also tried to capture long-range temporal and spatial dependencies in trajectories using attention mechanism \cite{lian2020geography,cui2021st,luo2021stan}.

Although considerable successes have been achieved by existing methods, next location recommendation is still an arduous problem. Mobility patterns and user preferences were two critical and interrelated factors in user-POI interaction modeling. Mobility patterns depict short-term travel states reflected by the recent check-ins of the user, which provide a more concrete spatial-temporal context, while user preferences are affected by long-term user behaviors and general tastes. To be more specific, mobility pattern tells the model about the current check-in and history check-ins, while user preference tells the model about user’s tendency about next location. For example, if a user currently being at a restaurant at noon, the mobility pattern part (ie, history encoder) will capture this information along with the user’s previous check-ins. The user preference part (ie, preference generator) learns from check-in history that the user likes to exercise after lunch. Then the user preference part tells the model about this tendency. Combined with the information from mobility pattern part, the model generates the most likely result, gym. Mobility patterns and user preferences should be both sufficiently well captured from historical check-in sequences to facilitate next location recommendation. However, most existing works \cite{lim2020stp,sun2020go,lian2020geography,luo2021stan} merely use a single compound module to learn mobility patterns and user preferences implicitly. Taking STAN \cite{luo2021stan} as an example, it adopts a bi-layer attention module to aggregate spatial-temporal correlation among POIs and recalls the target POI with consideration of personalized item frequencies. Such a strategy jointly models mobility patterns and user preferences. We argue that these two factors shall be modeled explicitly and separately due to the following reasons: 1) They are essentially heterogeneous information, where mobility patterns are a group of ordered POIs and user preferences are unordered attributes. It is better to represent mobility patterns using a sequence of embeddings that contain temporal transitions, while a user's preference can be captured well by only a single embedding. 2) They serve different purposes in location recommendation. If in an extreme case that two users have the same check-ins but different preferences, a good POI recommendation model should be able to generate different outputs. Another case is, for users with different historical check-in sequences (\eg two workers living and working in different places), they may have the same preferences (\eg only commuting every day). However, existing models which modeling user preferences and mobility patterns implicitly may not capture this underlying pattern well, resulting in inaccurate results.

To address the aforementioned issue, we attempt to solve the next location recommendation problem by formulating a new point of view. Specifically, a transformer-based framework, called POIFormer, is designed for end-to-end next location recommendation based on disentangled mobility pattern and user preference modeling. POIFormer first utilizes a history encoder to learn spatial-temporal mobility patterns from user's check-in history. Then, a query generator is employed to model the user preferences via contrastive learning. More specifically, historical check-in sequences are augmented and fed into the query generator. By maximizing the similarity between original and augmented views of the same check-in sequence, user-specific intention queries can be learned. These queries encode user's intention, while mobility patterns capture the current travel state. Then, the user preferences are regarded as queries, and the mobility patterns are keys and values. Preference decoder is then designed to decode user's intention and recommend next location. It should be noted that we also directly apply contrastive learning on inputs to model mobility pattern and user preference together, which leads to about $2\%\sim3\%$ performance degradation on \texttt{Recall@5} and \texttt{Recall@10} compared with POIFormer. The encoder in POIFormer can not only be a transformer, but also use other feature extraction modules. However, with merely using simple transformer layers, POIFormer has already achieved good results, which shows the superiority of our framework. Overall, the main contributions of this work can be summarized as follows:

\begin{figure*}
\centering
\includegraphics[width=0.9\textwidth]{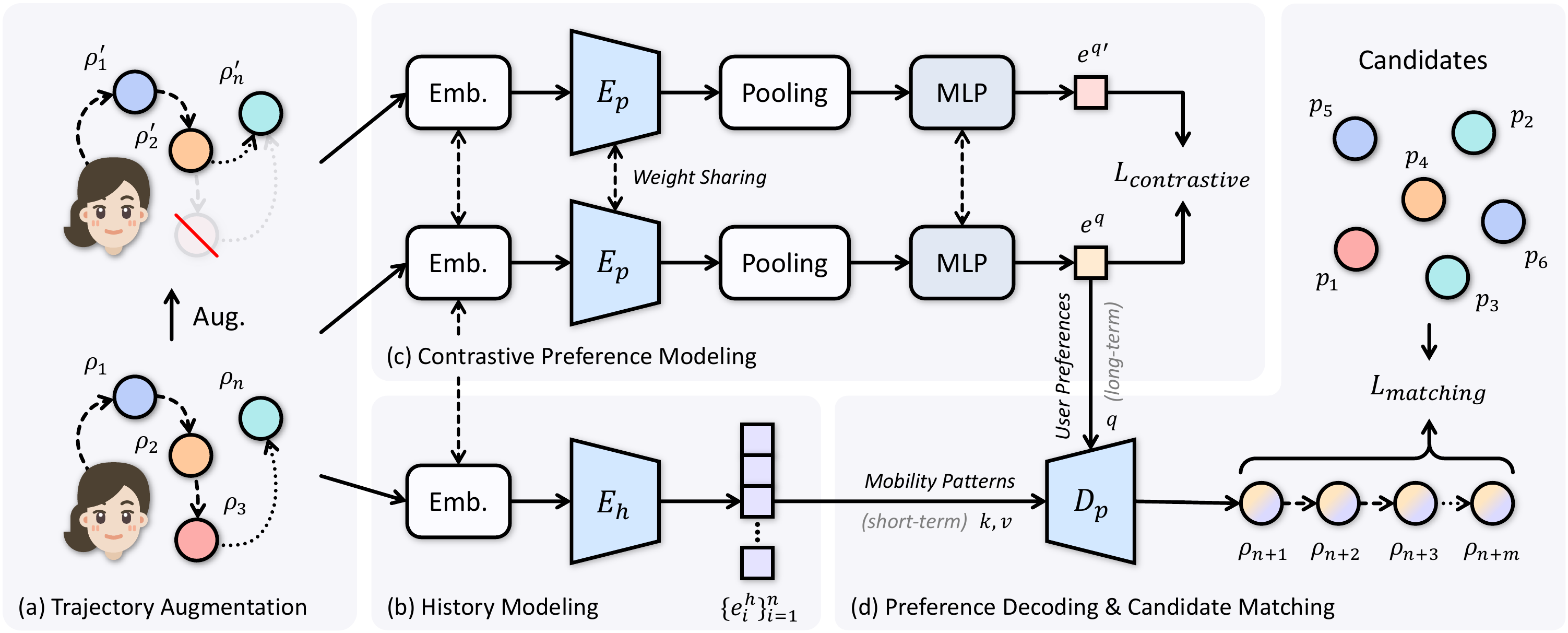}
\caption{The overall architecture of POIFormer. Historical check-in sequences $C = \{\rho_i\}_{i=1}^n$ are processed by history encoder $E_h$ and query generator $E_p$ to obtain mobility patterns $\{e_i^h\}_{i=1}^n$ and user preferences $e^q$. Contrastive learning is leveraged to enhance $e^q$ during training. Preference decoder $D_p$ finally generates personalized next location prediction $\{\rho_{n+i}\}_{i=1}^m$.}
\label{fig:2}
\end{figure*}

\begin{itemize}
\item We formulate a new point of view on next location recommendation. That is, mobility patterns and user preferences ought to be modeled explicitly and separately from historical check-in sequences.
\item A novel framework POIFormer is designed to realize disentangled mobility pattern and user preference modeling. A query generator is proposed for explicit user preference modeling by contrastive learning.
\item Experimental comparisons have been carried out on four real-world datasets, including Gowalla, Foursquare-SIN, Foursquare-TKY, and Brightkite, showing that the proposed scheme outperforms state-of-the-art counterparts under multiple settings.
\end{itemize}

\section{Related Work}

In this section, we extensively review the development of sequential recommendation, followed by the recent progress of next location recommendation. Some advances in contrastive learning and its applications are also considered.

\subsection{Sequential Recommendation}

Sequential recommendation is defined as a task that recommends the next item based on a sequence of existing items. Such a task has attracted lots of attention in recent years due to its great business value in multiple areas. The majority of sequential recommendation methods can be divided into Markov-based \cite{koren2009matrix,rendle2010factorization,cheng2013where,he2016fusing} and deep learning-based \cite{yu2016dynamic,li2017neural,massimo2018harnessing,zhou2019deep,yuan2019simple,wang2019towards,wu2019session,yu2020tagnn,kang2018self,li2020time,ye2020time} ones, in which Markov-based models focus on the transition matrix that determines the next behavior iteratively, while deep learning-based models handle the recommendation process with data-driven manners. Specifically, \cite{koren2009matrix} and \cite{rendle2010factorization} proposed to tackle the non-adjacent visits in sequences, which can not be easily modeled by Markov-based models. Other methods \cite{cheng2013where,he2016fusing} tried to incorporate spatial-temporal information into existing models. With the help of recurrent neural networks (RNNs), methods for session-based recommendation \cite{li2017neural}, next basket recommendation \cite{yu2016dynamic} and next item recommendation \cite{zhou2019deep} achieved considerable performances. Some other approaches also leverage convolutional neural networks (CNNs) \cite{yuan2019simple,wang2019towards}, graph neural networks (GNNs) \cite{wu2019session,yu2020tagnn}, and reinforcement learning (RL) \cite{massimo2018harnessing} to realize effective sequential recommendation. More recently, with the development of transformers \cite{vaswani2017attention}, self-attention mechanism is widely used in current approaches \cite{kang2018self,li2020time,ye2020time}.

\subsection{Next Location Recommendation}

Next location recommendation, also called next POI recommendation, is a typical downstream variant of sequential recommendation. It requires that the model can learn not only the users' mobility patterns, but also the spatial-temporal information among check-ins and locations. Similarly, early deep learning-based location recommendation approaches are built upon RNN and its variants \cite{liu2016predicting,wu2016sape,yao2017serm,zhu2017next,feng2018deepmove,huang2019attention,zhao2020where,sun2020go}. Liu \etal \cite{liu2016predicting} proposed STRNN, which enhances the ability of RNNs in spatial-temporal modeling by adopting spatial and temporal intervals between consecutive check-ins. This strategy was also studied by many of the following works. For instance, Time-LSTM \cite{zhu2017next} added temporal information into long short-term memory networks (LSTM), and STGN \cite{zhao2020where} further incorporates spatial information by designing spatial-temporal gates. Attention mechanism \cite{huang2019attention,lian2020geography,luo2021stan} was also adopted later on. Here, STAN \cite{luo2021stan} and its baseline, GeoSAN \cite{lian2020geography}, utilized self-attention to model long-range dependencies in check-in sequences. They claim that distant check-ins may sometimes provide high-level semantic information revealing user behaviors. To the best of our knowledge, POIFormer is the first method that explicitly models user preferences.

\subsection{Contrastive Learning}

As a representative type of self-supervised learning, contrastive learning optimizes semantic representations by contrasting positive samples against negative samples. Such a learning scheme has been widely used in computer vision and natural language processing communities \cite{chen2020simple,he2020momentum,giorgi2020declutr,chen2021exploring}. Here, SimCLR \cite{chen2020simple} and MoCo \cite{he2020momentum} are two typical contrastive learning paradigms using both positive and negative samples. SimCLR introduces a simple but effective framework with two siamese encoders and projection heads to realize self-supervised representation learning. MoCo tackles the problem of large batch size by introducing memory bank, storing the hidden representations of negative samples. In our context, user preferences can be learned via contrastive learning on the historical trajectory. The original and augmented views of the same trajectory are regarded as positive samples, while the remaining ones are negative. Experimental comparisons demonstrate that such a contrastive user preference modeling paradigm can boost performances of next location recommendation.

\begin{figure*}
\centering
\includegraphics[width=0.975\textwidth]{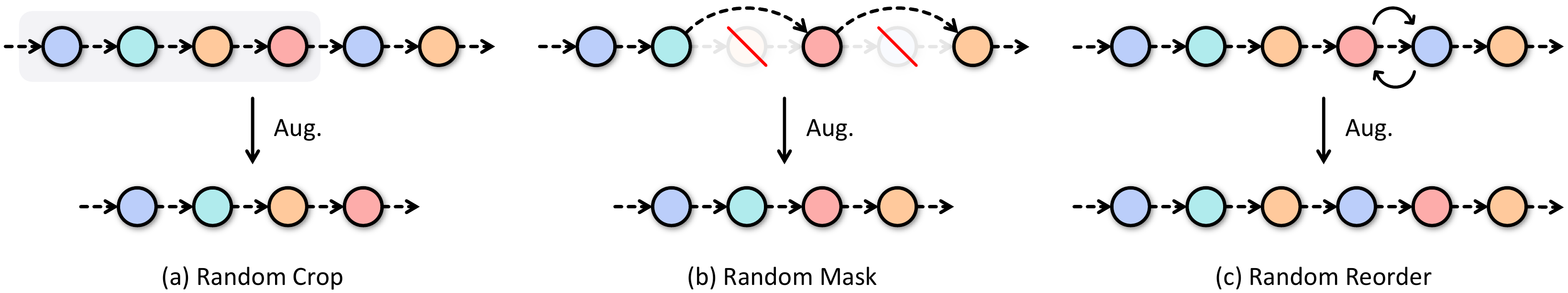}
\caption{Examples of three preference-irrelevant check-in augmentation approaches used in POIFormer.}
\label{fig:3}
\end{figure*}

\section{Method}

In this section, we introduce more details about the proposed POIFormer. We first expound on the problem statement and provide an overview of the framework. Then, we elaborate on the three main modules of POIFormer, \ie history encoder, query generator, and preference decoder.

\subsection{Overview}

A check-in $\rho_{t_i}^u = (u, t_i, p_i)$ denotes a behavior that a user $u$ visits POI $p_i$ at time $t_i$. Noticeably, each POI $p_i$ contains several meta information (\eg location and category). Each user $u$ is associated with a sequence of historical check-ins $C_{t_1 \rightarrow t_n}^u = \{\rho_{t_i}^u\}_{i=1}^n$, capturing the user's spatial information in temporal order. Given a check-in sequence, next location recommendation aims to predict the next POI $\rho_{t_{n+1}}$ that the user may visit. For simplicity, we deprecate the superscript $u$ and subscript $t_i$ in the following sections, that is, a check-in sequence can also be represented by $C = \{\rho_i\}_{i=1}^n$.

As can be seen from Figure~\ref{fig:2}, the overall architecture of our framework originates from the transformer encoder-decoder structure, and it can be divided into three parts, \ie history encoder, query generator, and preference decoder. The historical check-in sequence $C$ is firstly fed into the multi-modal embedding module to generate hidden representations $\{e_i^\rho\}_{i=1}^n$. These representations are then processed by history encoder to obtain mobility pattern representations $\{e_i^h\}_{i=1}^n$, and query generator to learn user-specific intention query $e^q$ by user preference modeling via contrastive learning. Subsequently, mobility patterns $\{e_i^h\}_{i=1}^n$ and the intention query $e^q$ are adopted as memory and query for preference decoder, which is designed to make personalized predictions for the next locations.

\subsection{Multi-modal Embedding}

Unlike applications in computer vision and natural language processing, properties of check-ins are naturally discrete and heterogeneous, and thus can not be directly regarded as inputs for models. Therefore, a multi-modal embedding module is necessary to embed check-ins by encoding time, location, and POI category into continuous latent representations. In greater detail, we discretize the timestamps into the hours in a week, denoted by $24 \times 7 = 168$-d one-hot vectors. These vectors are then linearly projected into $d$-dimensional temporal embeddings $e^t \in \mathbb{R}^d$. Location embeddings $e^l \in \mathbb{R}^d$ can similarly be obtained by mapping 2-d normalized longitude-latitude vectors using linear projections. For POI categories, we adopt the projected GloVe \cite{pennington2014glove} embeddings of category labels $e^c \in \mathbb{R}^d$. Therefore, the embedding of each POI can be regarded as the sum of location and category embeddings $e_i^p = e_i^l + e_i^c$, and the embedding of each check-in also include the temporal information as $e_i^\rho = e_i^p + e_i^t$. We also add a learnable positional embedding to the multi-modal embedding for modeling the order of successive POIs in a sequence.

\subsection{History Encoder}

To capture the spatial-temporal dependencies in historical check-in sequences, we construct the history encoder by stacking transformer encoder layers \cite{vaswani2017attention}, each consisting of a multi-head self-attention module and a feed-forward network. Residual connections and layer normalizations are also employed. In each attention head, self-attention for check-ins $\{\rho_i\}_{i=1}^n$ can be formulated as
\begin{gather}\label{eq1}
e_i^{\rho\prime} = e_i^\rho + w_z \sum_{j=1}^n \frac{\exp(w_q e_i^\rho \times w_k e_j^\rho)}{\sum_{m=1}^n \exp(w_q e_i^\rho \times w_k e_m^\rho)} w_v e_j^\rho
\end{gather}
where $e_i^\rho$ and $e_i^{\rho\prime}$ are the input and output embeddings of $\rho_i$, and $w_{\{q, k, v, z\}}$ denotes linear transform weights for the query, key, value, and output matrices. Self-attention mechanism aggregates global context information into each check-in embedding, so that history encoder is able to jointly attend to information from different embedding sub-spaces at different spatial-temporal positions. After aggregating the features, a two-layer feed-forward network formed by \texttt{Linear\,$\rightarrow$\,ReLU\,$\rightarrow$\,Dropout\,$\rightarrow$\,Linear} is used to further project the check-in embeddings.

\subsection{Query Generator}

Inspired by the success of contrastive learning in computer vision \cite{chen2020simple} and natural language processing \cite{gao2021simcse}, we attempt to apply this self-supervised paradigm to next location recommendation. The core of contrastive learning is to leverage self-supervised signals derived from unlabeled raw data to enhance the model performances. By maximizing the similarity between differently augmented views of the same sample, more distinguishable representations can be learned. In our case, check-in sequences and their augmented ones can be regarded as positive pairs, while other combinations should be negative pairs. Query generator consists of three parts, including a trajectory augmentation module, a preference encoder, and a contrastive learning objective.

\subsubsection{Trajectory Augmentation}

Stochastic trajectory augmentation is utilized to randomly transform check-in sequences into other views, in which the implicit user preferences remain unchanged. The three preference-irrelevant augmentation methods we use are illustrated in Figure~\ref{fig:3}, and they are discussed in detail as follows.

\begin{itemize}
\item \textbf{Random Crop:} Random crop mains cropping continuous sub-sequences of check-ins from original views. It provides a local perspective of the users' historical behaviors, forcing the model to learn more generalized user preference representations without comprehensive information of the whole check-in sequences.
\item \textbf{Random Mask:} Random mask can also be called as temporal dropout. It randomly masks out a small subset of non-neighboring check-ins. Masked sequences still reserve the main preference information of the users, enhancing the robustness of learned representations. An intuitive example is that if a user always goes to gym after lunch, removing one gym check-in will not affect the learning of preference embedding.
\item \textbf{Random Reorder:} Random reorder indicates randomly interchanging the order of two neighboring check-ins. Such a perturbation can push the model to rely less on check-in orders and learn temporal irrelevant preferences rather than simple mobility patterns. For example, it will also not affect preference representation whether a student goes to the snack shop first or print shop first.
\end{itemize}

The augmentations above are also considered in previous works \cite{xie2020contrastive} for sequential item recommendation. In practice, we randomly sample a method and apply it to a user's historical check-in sequence. Each check-in sequence, together with its augmented view, is considered as a positive pair. Unpaired sequences and augmented views in the same batch are regarded as negative pairs.

\subsubsection{Preference Encoder}

Preference encoders are employed to extract temporal irrelevant user preferences from check-in sequences. Here, the choice of preference encoder architecture is flexible. we adopt the same transformer encoder \cite{vaswani2017attention} as history encoder for simplicity. The model parameters are shared between the two encoders as well as projection heads for original views and augmented views, respectively. After obtaining the pooled and projected representations of original and augmented check-in sequences, \ie $e^q$ and $e^{q\prime}$, a contrastive learning objective \cite{oord2018representation} is applied to maximize the mutual information of positive pairs.

\subsubsection{Contrastive Learning Objective}

In order to enhance the user preference representations, an InfoNCE loss \cite{oord2018representation} is applied to realize contrastive prediction that can identify positive samples for a given sequence. This loss aims to maximize the agreement between the original and augmented views of the same historical check-in sequence, and minimize the agreement between unpaired views. The formulation of contrastive loss is
\begin{gather}
\alpha_i = -\frac{1}{2}\log\frac{\exp(e_i^q \cdot e_i^{q\prime}/\tau)}{\sum_{j=1}^B\exp(e_i^q \cdot e_j^{q\prime}/\tau)} \label{eq2} \\
\beta_i = -\frac{1}{2}\log\frac{\exp(e_i^{q\prime} \cdot e_i^q /\tau)}{\sum_{j=1}^B\exp(e_i^{q\prime} \cdot e_j^q/\tau)} \label{eq3} \\
\mathcal{L}_{contrastive} = \frac{1}{B}\sum_{i=1}^B(\alpha_i + \beta_i)
\end{gather}
where $B$ denotes the batch size, $\{e_i^q, e_i^{q\prime}\}$ means a pair of original and augmented views of a check-in sequence, and $\tau$ is a learnable temperature parameter controlling the smoothness of normalization. This loss can be regarded as an average of two $B$-way cross-entropy losses in a batch.

\begin{table}
\small
\setlength\tabcolsep{0pt}
\caption{Statistics of the datasets in our experiments.}
\label{tab:1}
\begin{tabularx}{\linewidth}{@{\hspace{2.35mm}}p{1.7cm}<{\raggedright}|p{1.625cm}<{\centering}p{1.625cm}<{\centering}p{1.625cm}<{\centering}p{1.625cm}<{\centering}}
\toprule
& Gowalla & SIN & TKY & Brightkite \\
\midrule
\#Users & 53,008 & 2,032 & 2,245 & 51,406 \\
\#Locations & 121,944 & 3,662 & 7,872 & 772,967 \\
\#Check-ins & 3,302,414 & 179,721 & 447,571 & 4,747,288 \\
\bottomrule
\end{tabularx}
\end{table}

\subsection{Preference Decoder}

Preference decoder takes user intention queries (\ie user preference representations) $e^q$ and mobility pattern representations $\{e_i^h\}_{i=1}^n$ as inputs, and predicts the personalized next locations. Specifically, it derives from the standard transformer decoder \cite{vaswani2017attention} which contains several stackable blocks, each consisting of a multi-head self-attention module, a multi-head cross-attention module, and a feed-forward network. Mobility patterns and user preferences are deeply fused in these blocks. In each attention head, cross-attention can be computed by
\begin{gather}
e^p = e^{q\prime} + w_z \sum_{j=1}^n \frac{\exp(w_q e^{q\prime} \times w_k e_j^h)}{\sum_{m=1}^n \exp(w_q e^{q\prime} \times w_k e_m^h)} w_v e_j^\rho
\end{gather}
where $e^{q\prime}$ is the self-attention output, and other notations are consistent with Eq.~\ref{eq1}. The output $e^p$ is then fed into the feed-forward network to obtain $e^{p\prime}$, which serves as the input query of the next block. Embedding of the predicted next POI $\hat{e}_{n+1}^p$ should be the output of the final block. Note that the query can also be shifted right iteratively to estimate the embedding of k-hop next POI $\hat{e}_{n+k}^p$.

\begin{table*}
\small
\setlength\tabcolsep{0pt}
\caption{Comparison with representative methods on multiple datasets.}
\label{tab:2}
\begin{tabularx}{\linewidth}{@{\hspace{1.5mm}}p{4.35cm}<{\raggedright}p{1.6cm}<{\centering}p{1.6cm}<{\centering}p{1.25mm}<{\centering}p{1.6cm}<{\centering}p{1.6cm}<{\centering}p{1.25mm}<{\centering}p{1.6cm}<{\centering}p{1.6cm}<{\centering}p{1.25mm}<{\centering}p{1.6cm}<{\centering}p{1.6cm}<{\centering}}
\toprule
\multirow{2.65}{4.3cm}{\centering Method} & \multicolumn{2}{c}{Gowalla} && \multicolumn{2}{c}{SIN} && \multicolumn{2}{c}{TKY} && \multicolumn{2}{c}{Brightkite} \\
\cmidrule{2-3} \cmidrule{5-6} \cmidrule{8-9} \cmidrule{11-12}
& Recall@5 & Recall@10 && Recall@5 & Recall@10 && Recall@5 & Recall@10 && Recall@5 & Recall@10 \\
\midrule
GRU \cite{cho2014learning} & 0.164 & 0.251 && 0.182 & 0.275 && 0.186 & 0.209 && 0.401 & 0.438 \\
STRNN \cite{liu2016predicting} & 0.166 & 0.257 && 0.184 & 0.279 && 0.179 & 0.202 && 0.423 & 0.448 \\
HST-LSTM \cite{kong2018hst} & 0.179 & 0.265 && 0.193 & 0.302 && 0.194 & 0.257 && 0.468 & 0.504 \\
DeepMove \cite{feng2018deepmove} & 0.196 & 0.270 && 0.268 & 0.351 && 0.239 & 0.316 && 0.503 & 0.587 \\
ATST-LSTM \cite{huang2019attention} & 0.212 & 0.278 && 0.275 & 0.359 && 0.251 & 0.330 && 0.517 & 0.593 \\
STGN \cite{zhao2020where} & 0.153 & 0.242 && 0.194 & 0.271 && 0.229 & 0.273 && 0.481 & 0.504 \\
ARNN \cite{guo2020attentional} & 0.181 & 0.275 && 0.185 & 0.270 && 0.182 & 0.254 && 0.429 & 0.518 \\
LSTPM \cite{sun2020go} & 0.202 & 0.270 && 0.257 & 0.331 && 0.258 & 0.333 && 0.438 & 0.524 \\
GeoSAN \cite{lian2020geography} & 0.276 & 0.365 && 0.296 & 0.374 && 0.340 & 0.394 && 0.558 & 0.672 \\
STAN \cite{luo2021stan} & 0.302 & 0.400 && 0.346 & 0.426 && 0.375 & 0.430 && 0.569 & 0.673 \\
\midrule
\textbf{POIFormer} (Ours) & \textbf{0.328} & \textbf{0.433} && \textbf{0.406} & \textbf{0.469} && \textbf{0.382} & \textbf{0.470} && \textbf{0.576} & \textbf{0.682} \\
\bottomrule
\end{tabularx}
\end{table*}

\begin{table*}
\small
\setlength\tabcolsep{0pt}
\caption{Effectiveness justification of different modules in POIFormer.}
\label{tab:3}
\begin{tabularx}{\linewidth}{p{1.425cm}<{\centering}p{1.425cm}<{\centering}p{1.425cm}<{\centering}p{2.25mm}<{\centering}p{1.6cm}<{\centering}p{1.6cm}<{\centering}p{1.25mm}<{\centering}p{1.6cm}<{\centering}p{1.6cm}<{\centering}p{1.25mm}<{\centering}p{1.6cm}<{\centering}p{1.6cm}<{\centering}p{1.25mm}<{\centering}p{1.6cm}<{\centering}p{1.6cm}<{\centering}}
\toprule
\multirow{2.55}{1.45cm}{\centering History Encoder} & \multirow{2.55}{1.45cm}{\centering Query Generator} & \multirow{2.55}{1.45cm}{\centering Contrastive Loss} && \multicolumn{2}{c}{Gowalla} && \multicolumn{2}{c}{SIN} && \multicolumn{2}{c}{TKY} && \multicolumn{2}{c}{Brightkite} \\
\cmidrule{5-6} \cmidrule{8-9} \cmidrule{11-12} \cmidrule{14-15}
&&&& Recall@5 & Recall@10 && Recall@5 & Recall@10 && Recall@5 & Recall@10 && Recall@5 & Recall@10 \\
\midrule
\checkmark &&&& 0.306 & 0.412 && 0.347 & 0.429 && 0.364 & 0.435 && 0.552 & 0.657 \\
&& \checkmark && 0.313 & 0.418 && 0.354 & 0.441 && 0.368 & 0.443 && 0.559 & 0.660 \\
\checkmark & \checkmark &&& 0.320 & 0.421 && 0.393 & 0.457 && 0.374 & 0.459 && 0.573 & 0.675 \\
\checkmark & \checkmark & \checkmark && \textbf{0.328} & \textbf{0.433} && \textbf{0.406} & \textbf{0.469} && \textbf{0.382} & \textbf{0.470} && \textbf{0.576} & \textbf{0.682} \\
\bottomrule
\end{tabularx}
\end{table*}

\subsection{Model Training \& Inference}

We compute the cosine similarities between the embeddings of the predicted POI $\hat{e}_{n+1}^p$ and candidates $\{e_i^p\}_{i=1}^N$. Here, $N$ is the number of candidates. The candidate POI with the largest similarity with the predicted next POI is regarded as the model output. During training, we compute the cross-entropy among the paired candidate and sampled unpaired candidates, and adopt it as the matching loss:
\begin{gather}
\mathcal{L}_{matching} = -\log\frac{\exp(\hat{e}_{n+1}^p \cdot e_0^p/\tau)}{\sum_{i=0}^{N_s}\exp(\hat{e}_{n+1}^p \cdot e_i^p/\tau)}
\end{gather}
where $e_0^p$ is the embedding of the positive candidate, and $N_s$ denotes the number of negative samples. Other notations are consistent with 	Eq.~\ref{eq2} and Eq.~\ref{eq3}. This POI matching objective is jointly optimized with the aforementioned contrastive one, so that the final total loss can be formulated as:
\begin{gather}
\mathcal{L}_{total} = \mathcal{L}_{matching} + \lambda\mathcal{L}_{contrastive}
\end{gather}
where $\lambda$ is a re-weighting term for balancing the two losses. We regard candidate POI matching as the main task and contrastive preference modeling as the auxiliary task. Thus, the model can also be trained without $\mathcal{L}_{contrastive}$. Detailed ablation studies are reported in the following section.

\section{Experiments}

In this section, we report the results of the extensive experiments to evaluate the performance and show the superiority of our method. The results also demonstrate the effectiveness and utility of POIFormer. Further experiments also validate the rationality of each component of POIFormer.

\subsection{Experimental Settings}

\subsubsection{Datasets}

We leverage four public real-world datasets, \ie Gowalla, SIN, TKY, and Brightkite to evaluate the performances of POIFormer and its counterparts. The last three datasets were collected in Singapore, Tokyo, and New York from Foursquare, respectively. Some statistical details of each dataset are illustrated in Table~\ref{tab:1}. Please note that these four datasets are benchmark datasets in POI recommendation \cite{luo2021stan,xue2021mobtcast}.

To make the results directly comparable, we follow the data pre-processing strategy of existing work that filters out low-frequency users and POIs and uses sliced trajectories with a fixed length window (\ie 100 check-ins) to filter out extreme long sequences. We also adopt the same \texttt{train/val/test} split as STAN \cite{luo2021stan}. Specifically, for each user, the last check-in is held out as the test data, and the check-in just before the last one is for validation. The remainings are used for training. It is worth noting that we also make full use of history sequences by slicing them into sub-sequences during training. For instance, assume a user has $n$ check-ins in the historical sequence, the length of training set is $n - 3$, with the first $[1, n - 3]$ check-ins as inputs and the $[2, n - 2]$ visited locations as labels.

\begin{table*}
\small
\setlength\tabcolsep{0pt}
\caption{Effectiveness justification of different augmentation methods in POIFormer.}
\label{tab:4}
\begin{tabularx}{\linewidth}{@{\hspace{1mm}}p{1.425cm}<{\centering}p{1.425cm}<{\centering}p{1.425cm}<{\centering}p{1.25mm}<{\centering}p{1.6cm}<{\centering}p{1.6cm}<{\centering}p{1.25mm}<{\centering}p{1.6cm}<{\centering}p{1.6cm}<{\centering}p{1.25mm}<{\centering}p{1.6cm}<{\centering}p{1.6cm}<{\centering}p{1.25mm}<{\centering}p{1.6cm}<{\centering}p{1.6cm}<{\centering}}
\toprule
\multirow{2.55}{1.45cm}{\centering Random Crop} & \multirow{2.55}{1.45cm}{\centering Random Mask} & \multirow{2.55}{1.45cm}{\centering Random Reorder} && \multicolumn{2}{c}{Gowalla} && \multicolumn{2}{c}{SIN} && \multicolumn{2}{c}{TKY} && \multicolumn{2}{c}{Brightkite} \\
\cmidrule{5-6} \cmidrule{8-9} \cmidrule{11-12} \cmidrule{14-15}
&&&& Recall@5 & Recall@10 && Recall@5 & Recall@10 && Recall@5 & Recall@10 && Recall@5 & Recall@10 \\
\midrule
&&&& 0.320 & 0.421 && 0.393 & 0.457 && 0.374 & 0.459 && 0.573 & 0.675 \\
\checkmark &&&& 0.322 & 0.426 && 0.397 & 0.459 && 0.373 & 0.461 && 0.562 & 0.667 \\
& \checkmark &&& 0.324 & 0.428 && 0.401 & 0.464 && 0.372 & 0.460 && 0.570 & 0.673 \\
&& \checkmark && 0.319 & 0.422 && 0.394 & 0.457 && 0.366 & 0.452 && 0.568 & 0.669 \\
\midrule
\checkmark & \checkmark &&& 0.325 & 0.431 && \textbf{0.408} & 0.466 && 0.379 & 0.468 && 0.571 & 0.678 \\
\checkmark && \checkmark && 0.321 & 0.428 && 0.402 & 0.463 && 0.380 & \textbf{0.471} && 0.569 & 0.674 \\
& \checkmark & \checkmark && 0.324 & 0.426 && 0.399 & 0.462 && 0.372 & 0.463 && 0.572 & 0.681 \\
\midrule
\checkmark & \checkmark & \checkmark && \textbf{0.328} & \textbf{0.433} && 0.406 & \textbf{0.469} && \textbf{0.382} & 0.470 && \textbf{0.576} & \textbf{0.682} \\
\bottomrule
\end{tabularx}
\end{table*}

\subsubsection{Evaluation Metrics}

Following previous works \cite{luo2021stan,xue2021mobtcast}, we utilize recall values to evaluate the model performances. Here, \texttt{R@k} is defined as the proportion of true positive samples in all positive samples, while only the first $k$ model outputs are considered. Since there is only one ground truth for each check-in sequence, so that recall and accuracy are mathematically equivalent. Other evaluation metrics (\eg NDCG) and results on other datasets are reported in the appendix.

\subsubsection{Implementation Details}

In all experiments, each model consists of three history encoder, query generator, and preference decoder layers each, unless noted otherwise. We set the hidden dimensions to $256$, with $4\times$ dimension expansions in FFNs. Learnable positional encodings, pre-norm style layer normalizations \cite{ba2016layer}, $8$ attention heads, and $0.1$ dropout rates are adopted in all encoder and decoder layers. The temperature parameter is set as $\tau = 1.0$, and all the weights of losses are set to $1.0$. The probabilities of augmentations are set to $0.7$. During training, we use Adam optimizer \cite{kingma2015adam} with $1e$$-$$3$ learning rate and 1e-4 weight decay. The models are trained for $100$ epochs with batch size $16$.

For each baseline method, all other hyper-parameters follow the suggestions from the original settings in their papers. The reported performances of baselines are obtained under their optimal settings. For the purpose of fair comparison, we do not use any semantic information to construct knowledge graph in ARNN \cite{guo2020attentional}, as semantic information is not introduced by other methods. Details about the baselines are listed in the the appendix.

\subsection{Overall Performance Comparison}

We compare the performance of our method with the baselines mentioned above. The results are shown in Table~\ref{tab:2}. To make our experiment more convincing, we use a T-test with a $p$-value of $0.01$ to examine the performance improvement. It is demonstrated that the improvement of POIFormer is statistically significant. From the table, we can observe that:

Firstly, POIFormer consistently outperforms all compared representative methods by $8.3\%$ to $13.6\%$ with respect to \texttt{R@5} and \texttt{R@10}. This demonstrates the effectiveness of our method in next location recommendation task. Different from existing approaches which still focus on modeling spatial and temporal information from historical check-ins while neglecting user preference, explicit and disentangled mobility patterns and user preferences modeling can greatly improve the model performances. In addition, the contrastive learning paradigm adopted in the query generator takes full advantage of self-supervised information from noisy data, which contributes to the enhancement of the user preference representations.

Secondly, compared with RNN-based approaches, self-attention based ones such as GeoSAN \cite{lian2020geography} and STAN \cite{luo2021stan} clearly achieve better performances. It is reasonable since attention is more powerful for capturing global contextual information in spatial-temporal check-in sequences. Among RNN-based models, LSTPM \cite{sun2020go} and DeepMove \cite{feng2018deepmove} have relatively better performances than others, which attributes to their consideration of short-term and long-term periodicity modeling. This makes up for the inherent defects of RNN to a certain extent. Among self-attention models, GeoSAN introduces a geography hierarchical gridding encoder, while STAN further utilizes spatial-temporal intervals within the check-in sequences for modeling non-consecutive visits and non-adjacent locations. The superiority of POIFormer is demonstrated since it can also be regarded as a type of attention-based methods.

\subsection{Ablation Study}

\subsubsection{Contrastive Preference Modeling}

We conduct an extensive ablation study on POIFormer to verify the effectiveness of each proposed module. The comparisons are shown in Table~\ref{tab:3}. The first row is the basic version of our model, with only a history encoder that jointly models mobility patterns and user preferences. The second row shows that directly adopting contrastive learning without query generator can only bring limited improvements. The third row indicates the model with query generator and preference decoder, its comparison with the first row shows that explicit and disentangled mobility patterns and user preferences modeling can indeed boost the performances. The last row is our full model, demonstrating that contrastive learning can further enhance the user preference representations. Note that the basic version of POIFormer is already highly competitive with existing methods. This is because: 1) the design of our multi-modal embedding module is simple but effective, converting sparse spatial-temporal information into dense representations, and 2) our model can naturally be trained in parallel, while the previous state-of-the-art \cite{luo2021stan} can only be trained iteratively. Their data pre-processing and training are extremely time and memory consuming, with more than $10\times$ and $50\times$ time and memory consumption, respectively.

\subsubsection{Trajectory Augmentation Methods}

We also study the impact of each trajectory augmentation method by running models with different augmentation combinations. The results are shown in Table~\ref{tab:4}. Compared with not using contrastive learning, adopting only one augmentation method brings slight contributions. When sampling one of two methods, the user preference representations can be better enhanced. The last row shows that stronger augmentations can further boost the performances. We can also observe that random crop and random mask contribute a bit more than random reorder, since they do not change the POI visiting order and can be easier to generalize.

\section{Conclusion}

In this paper, we formulated a new point of view on next location recommendation that mobility patterns and user preferences ought to be modeled explicitly and separately. From this perspective, a novel POIFormer has been proposed to achieve end-to-end next location recommendation with contrastive user preference modeling. The three components of POIFormer, namely, history encoder, query generator, and preference decoder have been proven to be effective for capturing disentangled mobility pattern and user preference information from historical check-in sequences. The effectiveness and superiority of the proposed framework have been demonstrated on several real-world datasets, in comparison with current existing schemes under various settings.

\bibliographystyle{ijcai}
\bibliography{main}

\begin{thebibliography}{}

\bibitem[\protect\citeauthoryear{Afzali \bgroup \em et al.\egroup
  }{2021}]{afzali2021pointrec}
Jafar Afzali, Aleksander~Mark Drzewiecki, and Krisztian Balog.
\newblock Pointrec: A test collection for narrative-driven point of interest
  recommendation.
\newblock In {\em SIGIR}, pages 2478--2484, 2021.

\bibitem[\protect\citeauthoryear{Ba \bgroup \em et al.\egroup
  }{2016}]{ba2016layer}
Jimmy~Lei Ba, Jamie~Ryan Kiros, and Geoffrey~E Hinton.
\newblock Layer normalization.
\newblock In {\em NeurIPS}, 2016.

\bibitem[\protect\citeauthoryear{Baumann \bgroup \em et al.\egroup
  }{2013}]{baumann2013influence}
Paul Baumann, Wilhelm Kleiminger, and Silvia Santini.
\newblock The influence of temporal and spatial features on the performance of
  next-place prediction algorithms.
\newblock In {\em UbiComp}, pages 449--458, 2013.

\bibitem[\protect\citeauthoryear{Chen and He}{2021}]{chen2021exploring}
Xinlei Chen and Kaiming He.
\newblock Exploring simple siamese representation learning.
\newblock In {\em CVPR}, pages 15750--15758, 2021.

\bibitem[\protect\citeauthoryear{Chen \bgroup \em et al.\egroup
  }{2020}]{chen2020simple}
Ting Chen, Simon Kornblith, Mohammad Norouzi, and Geoffrey Hinton.
\newblock A simple framework for contrastive learning of visual
  representations.
\newblock In {\em ICML}, pages 1597--1607, 2020.

\bibitem[\protect\citeauthoryear{Cheng \bgroup \em et al.\egroup
  }{2013}]{cheng2013where}
Chen Cheng, Haiqin Yang, Michael~R Lyu, and Irwin King.
\newblock Where you like to go next: Successive point-of-interest
  recommendation.
\newblock In {\em IJCAI}, pages 2605--–2611, 2013.

\bibitem[\protect\citeauthoryear{Cho \bgroup \em et al.\egroup
  }{2014}]{cho2014learning}
Kyunghyun Cho, Bart Van~Merriënboer, Caglar Gulcehre, Dzmitry Bahdanau, Fethi
  Bougares, Holger Schwenk, and Yoshua Bengio.
\newblock Learning phrase representations using rnn encoder-decoder for
  statistical machine translation.
\newblock In {\em EMNLP}, 2014.

\bibitem[\protect\citeauthoryear{Cui \bgroup \em et al.\egroup
  }{2021}]{cui2021st}
Qiang Cui, Chenrui Zhang, Yafeng Zhang, Jinpeng Wang, and Mingchen Cai.
\newblock St-pil: Spatial-temporal periodic interest learning for next
  point-of-interest recommendation.
\newblock In {\em CIKM}, pages 2960--2964, 2021.

\bibitem[\protect\citeauthoryear{Feng \bgroup \em et al.\egroup
  }{2018}]{feng2018deepmove}
Jie Feng, Yong Li, Chao Zhang, Funing Sun, Fanchao Meng, Ang Guo, and Depeng
  Jin.
\newblock Deepmove: Predicting human mobility with attentional recurrent
  networks.
\newblock In {\em WWW}, pages 1459--1468, 2018.

\bibitem[\protect\citeauthoryear{Gambs \bgroup \em et al.\egroup
  }{2012}]{gambs2012next}
Sébastien Gambs, Marc-Olivier Killijian, and Miguel~Núñez del Prado~Cortez.
\newblock Next place prediction using mobility markov chains.
\newblock In {\em MPM}, pages 1--6, 2012.

\bibitem[\protect\citeauthoryear{Gao \bgroup \em et al.\egroup
  }{2021}]{gao2021simcse}
Tianyu Gao, Xingcheng Yao, and Danqi Chen.
\newblock Simcse: Simple contrastive learning of sentence embeddings.
\newblock In {\em EMNLP}, pages 6894--6910, 2021.

\bibitem[\protect\citeauthoryear{Giorgi \bgroup \em et al.\egroup
  }{2021}]{giorgi2020declutr}
John Giorgi, Osvald Nitski, Bo~Wang, and Gary Bader.
\newblock Declutr: Deep contrastive learning for unsupervised textual
  representations.
\newblock In {\em ACL}, pages 879--895, 2021.

\bibitem[\protect\citeauthoryear{Guo \bgroup \em et al.\egroup
  }{2020}]{guo2020attentional}
Qing Guo, Zhu Sun, Jie Zhang, and Yin-Leng Theng.
\newblock An attentional recurrent neural network for personalized next
  location recommendation.
\newblock In {\em AAAI}, pages 83--90, 2020.

\bibitem[\protect\citeauthoryear{Hang \bgroup \em et al.\egroup
  }{2018}]{hang2018exploring}
Mengyue Hang, Ian Pytlarz, and Jennifer Neville.
\newblock Exploring student check-in behavior for improved point-of-interest
  prediction.
\newblock In {\em KDD}, pages 321--330, 2018.

\bibitem[\protect\citeauthoryear{He and McAuley}{2016}]{he2016fusing}
Ruining He and Julian McAuley.
\newblock Fusing similarity models with markov chains for sparse sequential
  recommendation.
\newblock In {\em ICDM}, pages 191--200, 2016.

\bibitem[\protect\citeauthoryear{He \bgroup \em et al.\egroup
  }{2020}]{he2020momentum}
Kaiming He, Haoqi Fan, Yuxin Wu, Saining Xie, and Ross Girshick.
\newblock Momentum contrast for unsupervised visual representation learning.
\newblock In {\em CVPR}, pages 9729--9738, 2020.

\bibitem[\protect\citeauthoryear{Huang \bgroup \em et al.\egroup
  }{2019}]{huang2019attention}
Liwei Huang, Yutao Ma, Shibo Wang, and Yanbo Liu.
\newblock An attention-based spatiotemporal lstm network for next poi
  recommendation.
\newblock {\em IEEE Transactions on Services Computing}, 14(6):1585--1597,
  2019.

\bibitem[\protect\citeauthoryear{Kang and McAuley}{2018}]{kang2018self}
Wang-Cheng Kang and Julian McAuley.
\newblock Self-attentive sequential recommendation.
\newblock In {\em ICDM}, pages 197--206, 2018.

\bibitem[\protect\citeauthoryear{Kingma and Ba}{2015}]{kingma2015adam}
Diederik~P Kingma and Jimmy Ba.
\newblock Adam: A method for stochastic optimization.
\newblock In {\em ICLR}, 2015.

\bibitem[\protect\citeauthoryear{Kong and Wu}{2018}]{kong2018hst}
Dejiang Kong and Fei Wu.
\newblock Hst-lstm: A hierarchical spatial-temporal long-short term memory
  network for location prediction.
\newblock In {\em IJCAI}, 2018.

\bibitem[\protect\citeauthoryear{Koren \bgroup \em et al.\egroup
  }{2009}]{koren2009matrix}
Yehuda Koren, Robert Bell, and Chris Volinsky.
\newblock Matrix factorization techniques for recommender systems.
\newblock {\em Computer}, 42(8):30--37, 2009.

\bibitem[\protect\citeauthoryear{Li \bgroup \em et al.\egroup
  }{2017}]{li2017neural}
Jing Li, Pengjie Ren, Zhumin Chen, Zhaochun Ren, Tao Lian, and Jun Ma.
\newblock Neural attentive session-based recommendation.
\newblock In {\em CIKM}, pages 1419--1428, 2017.

\bibitem[\protect\citeauthoryear{Li \bgroup \em et al.\egroup
  }{2018}]{li2018next}
Ranzhen Li, Yanyan Shen, and Yanmin Zhu.
\newblock Next point-of-interest recommendation with temporal and multi-level
  context attention.
\newblock In {\em ICDM}, pages 1110--1115, 2018.

\bibitem[\protect\citeauthoryear{Li \bgroup \em et al.\egroup
  }{2020}]{li2020time}
Jiacheng Li, Yujie Wang, and Julian McAuley.
\newblock Time interval aware self-attention for sequential recommendation.
\newblock In {\em WSDM}, pages 322--330, 2020.

\bibitem[\protect\citeauthoryear{Lian \bgroup \em et al.\egroup
  }{2020}]{lian2020geography}
Defu Lian, Yongji Wu, Yong Ge, Xing Xie, and Enhong Chen.
\newblock Geography-aware sequential location recommendation.
\newblock In {\em KDD}, pages 2009--2019, 2020.

\bibitem[\protect\citeauthoryear{Lim \bgroup \em et al.\egroup
  }{2020}]{lim2020stp}
Nicholas Lim, Bryan Hooi, See-Kiong Ng, Xueou Wang, Yong~Liang Goh, Renrong
  Weng, and Jagannadan Varadarajan.
\newblock Stp-udgat: Spatial-temporal-preference user dimensional graph
  attention network for next poi recommendation.
\newblock In {\em CIKM}, pages 845--854, 2020.

\bibitem[\protect\citeauthoryear{Liu \bgroup \em et al.\egroup
  }{2013}]{liu2013personalized}
Xin Liu, Yong Liu, Karl Aberer, and Chunyan Miao.
\newblock Personalized point-of-interest recommendation by mining users'
  preference transition.
\newblock In {\em CIKM}, pages 733--738, 2013.

\bibitem[\protect\citeauthoryear{Liu \bgroup \em et al.\egroup
  }{2016}]{liu2016predicting}
Qiang Liu, Shu Wu, Liang Wang, and Tieniu Tan.
\newblock Predicting the next location: A recurrent model with spatial and
  temporal contexts.
\newblock In {\em AAAI}, 2016.

\bibitem[\protect\citeauthoryear{Luo \bgroup \em et al.\egroup
  }{2021}]{luo2021stan}
Yingtao Luo, Qiang Liu, and Zhaocheng Liu.
\newblock Stan: Spatio-temporal attention network for next location
  recommendation.
\newblock In {\em WWW}, pages 2177--2185, 2021.

\bibitem[\protect\citeauthoryear{Massimo and
  Ricci}{2018}]{massimo2018harnessing}
David Massimo and Francesco Ricci.
\newblock Harnessing a generalised user behaviour model for next-poi
  recommendation.
\newblock In {\em RecSys}, pages 402--406, 2018.

\bibitem[\protect\citeauthoryear{Oord \bgroup \em et al.\egroup
  }{2018}]{oord2018representation}
Aaron van~den Oord, Yazhe Li, and Oriol Vinyals.
\newblock Representation learning with contrastive predictive coding.
\newblock Technical Report arXiv:1807.03748, 2018.

\bibitem[\protect\citeauthoryear{Pennington \bgroup \em et al.\egroup
  }{2014}]{pennington2014glove}
Jeffrey Pennington, Richard Socher, and Christopher~D Manning.
\newblock Glove: Global vectors for word representation.
\newblock In {\em EMNLP}, pages 1532--1543, 2014.

\bibitem[\protect\citeauthoryear{Rendle \bgroup \em et al.\egroup
  }{2010}]{rendle2010factorizing}
Steffen Rendle, Christoph Freudenthaler, and Lars Schmidt-Thieme.
\newblock Factorizing personalized markov chains for next-basket
  recommendation.
\newblock In {\em WWW}, pages 811--820, 2010.

\bibitem[\protect\citeauthoryear{Rendle}{2010}]{rendle2010factorization}
Steffen Rendle.
\newblock Factorization machines.
\newblock In {\em ICDM}, pages 995--1000, 2010.

\bibitem[\protect\citeauthoryear{Rumelhart \bgroup \em et al.\egroup
  }{1985}]{rumelhart1985learning}
David~E Rumelhart, Geoffrey~E Hinton, and Ronald~J Williams.
\newblock Learning internal representations by error propagation.
\newblock {\em Parallel Distributed Processing: Explorations in the
  Microstructure of Cognition: Foundations}, 1985.

\bibitem[\protect\citeauthoryear{Sun \bgroup \em et al.\egroup
  }{2020}]{sun2020go}
Ke~Sun, Tieyun Qian, Tong Chen, Yile Liang, Quoc Viet~Hung Nguyen, and Hongzhi
  Yin.
\newblock Where to go next: Modeling long- and short-term user preferences for
  point-of-interest recommendation.
\newblock In {\em AAAI}, pages 214--221, 2020.

\bibitem[\protect\citeauthoryear{Vaswani \bgroup \em et al.\egroup
  }{2017}]{vaswani2017attention}
Ashish Vaswani, Noam Shazeer, Niki Parmar, Jakob Uszkoreit, Llion Jones,
  Aidan~N Gomez, Lukasz Kaiser, and Illia Polosukhin.
\newblock Attention is all you need.
\newblock In {\em NeurIPS}, pages 5998--6008, 2017.

\bibitem[\protect\citeauthoryear{Wang \bgroup \em et al.\egroup
  }{2016}]{wang2022graph}
Zhaobo Wang, Yanmin Zhu, Qiaomei Zhang, Haobin Liu, Chunyang Wang, and Tong
  Liu.
\newblock Graph-enhanced spatial-temporal network for next poi recommendation.
\newblock {\em ACM Transactions on Knowledge Discovery from Data}, 16(6):1--21,
  2016.

\bibitem[\protect\citeauthoryear{Wang \bgroup \em et al.\egroup
  }{2018}]{wang2018tpm}
Weiqing Wang, Hongzhi Yin, Xingzhong Du, Quoc Viet~Hung Nguyen, and Xiaofang
  Zhou.
\newblock Tpm: A temporal personalized model for spatial item recommendation.
\newblock {\em ACM Transactions on Intelligent Systems and Technology},
  9(6):1--25, 2018.

\bibitem[\protect\citeauthoryear{Wang \bgroup \em et al.\egroup
  }{2019}]{wang2019towards}
Jingyi Wang, Qiang Liu, Zhaocheng Liu, and Shu Wu.
\newblock Towards accurate and interpretable sequential prediction: A cnn \&
  attention-based feature extractor.
\newblock In {\em CIKM}, pages 1703--1712, 2019.

\bibitem[\protect\citeauthoryear{Wu \bgroup \em et al.\egroup
  }{2016}]{wu2016sape}
Shu Wu, Qiang Liu, Ping Bai, Liang Wang, and Tieniu Tan.
\newblock Sape: A system for situation-aware public security evaluation.
\newblock In {\em AAAI}, pages 4401--4402, 2016.

\bibitem[\protect\citeauthoryear{Wu \bgroup \em et al.\egroup
  }{2019}]{wu2019session}
Shu Wu, Yuyuan Tang, Yanqiao Zhu, Liang Wang, Xing Xie, and Tieniu Tan.
\newblock Session-based recommendation with graph neural networks.
\newblock In {\em AAAI}, pages 346--353, 2019.

\bibitem[\protect\citeauthoryear{Xie \bgroup \em et al.\egroup
  }{2016}]{xie2016learning}
Min Xie, Hongzhi Yin, Hao Wang, Fanjiang Xu, Weitong Chen, and Sen Wang.
\newblock Learning graph-based poi embedding for location-based recommendation.
\newblock In {\em CIKM}, pages 15--24, 2016.

\bibitem[\protect\citeauthoryear{Xie \bgroup \em et al.\egroup
  }{2020}]{xie2020contrastive}
Xu~Xie, Fei Sun, Zhaoyang Liu, Jinyang Gao, Bolin Ding, and Bin Cui.
\newblock Contrastive pre-training for sequential recommendation.
\newblock Technical Report arXiv:2010.14395, 2020.

\bibitem[\protect\citeauthoryear{Xue \bgroup \em et al.\egroup
  }{2021}]{xue2021mobtcast}
Hao Xue, Flora Salim, Yongli Ren, and Nuria Oliver.
\newblock Mobtcast: Leveraging auxiliary trajectory forecasting for human
  mobility prediction.
\newblock In {\em NeurIPS}, 2021.

\bibitem[\protect\citeauthoryear{Yao \bgroup \em et al.\egroup
  }{2017}]{yao2017serm}
Di~Yao, Chao Zhang, Jianhui Huang, and Jingping Bi.
\newblock Serm: A recurrent model for next location prediction in semantic
  trajectories.
\newblock In {\em CIKM}, pages 2411--2414, 2017.

\bibitem[\protect\citeauthoryear{Ye \bgroup \em et al.\egroup
  }{2020}]{ye2020time}
Wenwen Ye, Shuaiqiang Wang, Xu~Chen, Xuepeng Wang, Zheng Qin, and Dawei Yin.
\newblock Time matters: Sequential recommendation with complex temporal
  information.
\newblock In {\em SIGIR}, pages 1459--1468, 2020.

\bibitem[\protect\citeauthoryear{Yu \bgroup \em et al.\egroup
  }{2016}]{yu2016dynamic}
Feng Yu, Qiang Liu, Shu Wu, Liang Wang, and Tieniu Tan.
\newblock A dynamic recurrent model for next basket recommendation.
\newblock In {\em SIGIR}, pages 729--732, 2016.

\bibitem[\protect\citeauthoryear{Yu \bgroup \em et al.\egroup
  }{2020}]{yu2020tagnn}
Feng Yu, Yanqiao Zhu, Qiang Liu, Shu Wu, Liang Wang, and Tieniu Tan.
\newblock Tagnn: Target attentive graph neural networks for session-based
  recommendation.
\newblock In {\em SIGIR}, pages 1921--1924, 2020.

\bibitem[\protect\citeauthoryear{Yuan \bgroup \em et al.\egroup
  }{2019}]{yuan2019simple}
Fajie Yuan, Alexandros Karatzoglou, Ioannis Arapakis, Joemon~M Jose, and
  Xiangnan He.
\newblock A simple convolutional generative network for next item
  recommendation.
\newblock In {\em WSDM}, pages 582--590, 2019.

\bibitem[\protect\citeauthoryear{Zhang and
  Chow}{2015}]{zhang2015spatiotemporal}
Jia-Dong Zhang and Chi-Yin Chow.
\newblock Spatiotemporal sequential influence modeling for location
  recommendations: A gravity-based approach.
\newblock {\em ACM Transactions on Intelligent Systems and Technology},
  7(1):1--25, 2015.

\bibitem[\protect\citeauthoryear{Zhao \bgroup \em et al.\egroup
  }{2020}]{zhao2020where}
Pengpeng Zhao, Anjing Luo, Yanchi Liu, Fuzhen Zhuang, Jiajie Xu, Zhixu Li,
  Victor~S Sheng, and Xiaofang Zhou.
\newblock Where to go next: A spatio-temporal gated network for next poi
  recommendation.
\newblock {\em IEEE Transactions on Knowledge and Data Engineering},
  34(5):2512--2524, 2020.

\bibitem[\protect\citeauthoryear{Zhou \bgroup \em et al.\egroup
  }{2019}]{zhou2019deep}
Guorui Zhou, Na~Mou, Ying Fan, Qi~Pi, Weijie Bian, Chang Zhou, Xiaoqiang Zhu,
  and Kun Gai.
\newblock Deep interest evolution network for click-through rate prediction.
\newblock In {\em AAAI}, pages 5941--5948, 2019.

\bibitem[\protect\citeauthoryear{Zhu \bgroup \em et al.\egroup
  }{2017}]{zhu2017next}
Yu~Zhu, Hao Li, Yikang Liao, Beidou Wang, Ziyu Guan, Haifeng Liu, and Deng Cai.
\newblock What to do next: Modeling user behaviors by time-lstm.
\newblock In {\em IJCAI}, pages 3602--3608, 2017.

\end{thebibliography}

\vfill
\setcounter{section}{0}
\renewcommand{\thesection}{\Alph{section}}
\setcounter{equation}{0}
\renewcommand{\theequation}{\Alph{equation}}
\setcounter{table}{0}
\renewcommand{\thetable}{\Alph{table}}
\setcounter{figure}{0}
\renewcommand{\thefigure}{\Alph{figure}}

\begin{table*}
\small
\setlength\tabcolsep{0pt}
\caption{More comparisons with representative methods on multiple datasets.}
\label{tab:a}
\begin{tabularx}{\linewidth}{@{\hspace{1.5mm}}p{4.35cm}<{\raggedright}p{1.6cm}<{\centering}p{1.6cm}<{\centering}p{1.25mm}<{\centering}p{1.6cm}<{\centering}p{1.6cm}<{\centering}p{1.25mm}<{\centering}p{1.6cm}<{\centering}p{1.6cm}<{\centering}p{1.25mm}<{\centering}p{1.6cm}<{\centering}p{1.6cm}<{\centering}}
\toprule
\multirow{2.65}{4.35cm}{\centering Method} & \multicolumn{2}{c}{Brightkite} && \multicolumn{2}{c}{SIN} && \multicolumn{2}{c}{TKY} && \multicolumn{2}{c}{NYC} \\
\cmidrule{2-3} \cmidrule{5-6} \cmidrule{8-9} \cmidrule{11-12}
& NDCG@5 & NDCG@10 && NDCG@5 & NDCG@10 && NDCG@5 & NDCG@10 && NDCG@5 & NDCG@10 \\
\midrule
FPMC \cite{rendle2010factorizing} & 0.385 & 0.390 && 0.103 & 0.130 && 0.064 & 0.080 && 0.176 & 0.204 \\
RNN \cite{rumelhart1985learning} & 0.368 & 0.425 && 0.059 & 0.080 && 0.188 & 0.207 && 0.193 & 0.210 \\
HST-LSTM \cite{kong2018hst} & 0.373 & 0.413 && 0.112 & 0.134 && 0.221 & 0.243 && 0.197 & 0.214 \\
DeepMove \cite{feng2018deepmove} & 0.391 & 0.432 && 0.118 & 0.134 && 0.251 & 0.272 && 0.216 & 0.238 \\
LSTPM \cite{sun2020go} & 0.349 & 0.385 && 0.092 & 0.120 && 0.282 & 0.309 && 0.161 & 0.186 \\
GeoSAN \cite{lian2020geography} & 0.423 & 0.460 && 0.118 & 0.159 && 0.484 & 0.518 && 0.208 & 0.253 \\
\midrule
\textbf{POIFormer} (Ours) & \textbf{0.459} & \textbf{0.493} && \textbf{0.154} & \textbf{0.201} && \textbf{0.517} & \textbf{0.542} && \textbf{0.245} & \textbf{0.297} \\
\bottomrule
\end{tabularx}
\end{table*}

\section*{Appendix}

In this document, we provide more detailed information about the baseline methods and experimental results to complement the main paper. Additional evaluation metrics are also incorporated to demonstrate the significance and effectiveness of the proposed method.

\section{Datasets}

\renewcommand{\footnotesize}{\scriptsize}
We compare the proposed POIFormer and its counterparts on five real-world datasets, \ie Gowalla\footnote{\url{https://snap.stanford.edu/data/loc-gowalla.html}}, Foursquare-SIN\footnote{\url{https://www.ntu.edu.sg/home/gaocong/data/poidata.zip}}, Foursquare-TKY\footnote{\url{https://www-public.imtbs-tsp.eu/~zhang_da/pub/dataset_tsmc2014.zip}}, Foursquare-NYC\textsuperscript{3}, and Brightkite\footnote{\url{http://snap.stanford.edu/data/loc-brightkite.html}}. The Foursquare datasets were collected in Singapore, Tokyo, and New York, respectively. We show the statistics of Foursquare-NYC dataset in Table~\ref{tab:b} to complement the main paper.
\renewcommand{\footnotesize}{\footnotesize}

\section{Evaluation Metrics}

Aside from recall, we also leverage NDCG as a metric to evaluate the performance of next location recommendation. NDCG is a metric that rewards the method ranks positive items in the first few positions of the top-$k$ ranking list. Similar to the main paper, we report $k=5$ and $k=10$ in our experiments as well.

\section{Baselines}

In order to verify the effectiveness of our proposed method, we compare it with the following representative next location recommendation baselines:

\begin{itemize}
\item \textbf{RNN} \cite{rumelhart1985learning}: a deep neural network specially designed for sequential modeling;
\item \textbf{GRU} \cite{liu2016predicting}: a computationally cheaper variant of LSTM for sequential modeling that serves as a strong baseline of deep learning based methods;
\item \textbf{FPMC} \cite{rendle2010factorizing}: a baseline model of sequence recommendation task in the early research period of trajectory prediction field;
\item \textbf{STRNN} \cite{liu2016predicting}: an RNN-based model that can handle local spatial-temporal contexts using time/distance-specific transition matrices for different time intervals/geographical distances;
\item \textbf{HST-LSTM} \cite{kong2018hst}: an LSTM-based model that incorporates spatial-temporal transfer factors with an encoder-decoder structure;
\item \textbf{DeepMove} \cite{feng2018deepmove}: a mobility prediction pipeline with a multi-modal embedding RNN for modeling human mobility factors, and a historical attention model for capturing multi-level periodicity;
\item \textbf{ATST-LSTM} \cite{huang2019attention}: an LSTM-based model that utilizes distance and time differences among trajectory points, and adopts the attention mechanism;
\item \textbf{STGN}  \cite{zhao2020where}: an LSTM-based model that introduces spatial-temporal gates for extracting spatial-temporal correlations between successive check-ins;
\item \textbf{ARNN} \cite{guo2020attentional}: a unified recurrent framework that jointly captures sequential regularity and transition regularities of neighbors via attention mechanism. The neighbors are modeled by a knowledge graph;
\item \textbf{LSTPM} \cite{sun2020go}: a method with a non-local network for learning long-term preference and a geo-dilated RNN for learning short-term preference;
\item \textbf{GeoSAN} \cite{lian2020geography}: a geography-aware model that utilizes hierarchical gridding for GPS information with a self-attention based geography encoder;
\item \textbf{STAN} \cite{luo2021stan}: a spatial-temporal attention network that explicitly aggregates all relevant check-ins in trajectories, not only just successive ones.
\end{itemize}

\begin{table}
\small
\centering
\caption{Statistics of the Foursquare-NYC dataset.}
\label{tab:b}
\begin{tabularx}{0.575\linewidth}{ccc}
\toprule
\#Users & \#Locations & \#Check-ins \\
\midrule
51,406 & 772,967 & 4,747,288 \\
\bottomrule
\end{tabularx}
\end{table}

\section{More Experimental Results}

The performance comparison on the new Foursquare-NYC dataset and using NDCG as evaluation metric is shown in Table~\ref{tab:a}. Compared with previous representative methods, our proposed POIFormer shows $8\%\sim9\%$ performance gains on multiple datasets. The comparison demonstrates the significance and effectiveness of the proposed scheme.

\end{document}